\documentclass[twocolumn,aps,floats,superscriptaddress,showpacs]{revtex4}
\usepackage{graphics,epsfig,amsfonts,amssymb,amsmath}
\begin{document}

\title{Anisotropic transport in the overdoped High-Tc superconductors
within the Van Hove scenario}

\author{Bel\'en Valenzuela}
\affiliation{Department of Physics, University of Guelph,
Ontario N1G 2W1, Canada}
\author{Mar\'{\i}a A. H. Vozmediano}
\affiliation{Unidad Asociada CSIC-UC3M, Departamento de Matem\'aticas,
Universidad Carlos III de Madrid,
E-28911 Legan\'es, Madrid, Spain}
%\author{Thomas P. Devereaux}
%\affiliation{
%Department of Physics, University of Waterloo, Ontario, N2L 3GI Canada.}

\begin{abstract}
Recent Raman experiments\cite{devereaux1}
done in $Bi_2 Sr_2 CaCu_2O_4$ have found that
antinodal quasiparticles
become insulating when the doping is varied from overdoped to
optimally doped while nodal quasiparticles remain metallic.
We propose a simple explanation based on the
incoherence of the spectral function in the antinodal direction
due to the strong scattering suffered by the quasiparticles
in the proximity of the Van Hove singularity.
\end{abstract}
\date{\today}
\maketitle

Recent Raman experiments\cite{devereaux1} have found a strong anisotropy
of the
electron relaxation in $Bi_2 Sr_2 CaCu_2O_4$ when the doping varies
from the optimal region to the overdoped region. While the nodal
quasiparticles remain metallic and hardly vary
with temperature and doping the antinodal quasiparticles
suffer a metal-insulator transition at $p\sim 0.16$. The result is found
in the overdoped region and at high temperatures, therefore it
cannot be explained with the pseudogap physics.

The Raman scattering has been proposed as
a complementary  technique to ARPES to
explore the {\bf k}-dependence of the quasiparticles
due to
its ability to select
different regions of the Brillouin zone
by changing the polarizations of incoming and outgoing
photons.
In the point group $D_{4h}$ of the square lattice,
there are three different Raman
channels: $B_{1g}$, $B_{2g}$ and $A_{1g}$.
The $B_{1g}$ channel projects out the antinodal region of the
Fermi surface, the $B_{2g}$ the nodal region, and the $A_{1g}$ is a
weighted average over the entire Brillouin zone. The
mentioned experiments refer to a strong anisotropy
in the Raman signal found from a comparison of
the doping and temperature dependence of the
$B_{1g}$ and $B_{2g}$ signals.

The Fermi surface of most hole-doped cuprates around
optimal doping is close to
a saddle point of the dispersion relation\cite{vhexp}.
 The possible relevance
of this fact to the superconducting transition as well
as to the anomalous behavior of the normal state was put
forward in the early times of the cuprates and gave rise to the
so-called Van Hove scenario\cite{vhscenario}. A renormalization
group analysis at zero temperature
using the two-patch model\cite{nuphysb97} around
the Van Hove points
showed that the electron self-energy has the form of the one
assumed in the marginal Fermi liquid\cite{marginal}.
As a consequence
the quasiparticle spectral weight $Z$ goes to zero
logarithmically at low energies.
This result was reproduced
using many-patch one-loop functional
renormalization group technique\cite{katanin} where it is also
shown that the scattering rate at high temperatures
is almost linear in frequency.

Renormalization group calculations also
show the pinning of the Fermi level to the Van Hove
singularity\cite{prl00}
%,kataninpinning
what implies that for filling fractions close and
below the Van Hove filling, the low energy properties of the
system will be dominated by it. These features
and the phase diagram
make of the VH model a sensible microscopic model to
address the low energy physics of the cuprates.

In reference \cite{devereaux3}
Devereaux establishes a relation
between the $B_{2g}$ signal with the
in-plane optical conductivity and between the $B_{1g}$ and the out of plane
optical conductivity at low frequencies.
It is shown that these similarities work
quite well in $Bi-2212$ and $Y-123$ but not in $La-214$. This last
discrepancy is attributed to charge ordering effects.
However the general tendency shows
that the in-plane momentum is at least partially conserved in
the c-axis transport. Around optimal filling the optical conductivity
has an incoherent behavior what matches with the result found
in Raman for the $B_{1g}$ signal (antinodal).

In ref. \cite{geli} a formalism was proposed
to study the suppression of interlayer tunneling by
inelastic processes in two dimensional systems in the clean limit.
A relation was established between departure
from Fermi liquid behavior
driven by electron correlations inside the layers and the out of plane
coherence. There it was shown that
the out of plane hopping  is always irrelevant if the Fermi level
of the interacting electrons lies at a Van Hove
singularity.

The renormalization group approach is a technique well suited
for weak coupling, therefore it might be appropriate to study the
optimal-overdoped
region where the effective interaction is weaker.
We propose as a possible explanation of the metal-insulator transition
found in the $B_{1g}$ signal (antinodal)
the incoherence of the transport when the
chemical potential of the planes is close to the Van Hove
filling. For filling fractions close and below
the singularity, the Van Hove filling is
an attractive point in the renormalization group sense and
the transport should be incoherent up
to the Van Hove filling.
Above the Van Hove filling the flow of the chemical potential
drives the system away of the singularity area and the system
recovers a Fermi liquid behavior.
Within this scenario we can also understand the incoherent behavior
of the c-axis optical conductivity in two different ways: either through
the relation between the Raman signal $B_{1g}$ and the c-axis optical
conductivity at low frequencies\cite{devereaux3},
or through the prediction of
incoherent transport when the chemical potential is at or
below the Van Hove filling\cite{geli}.

In reference\cite{devereaux1}  a phenomenological model
was proposed to explain the metal insulator transition
in the antinodal region by invoking the
opening of an anisotropic gap in the hot
spots region (Van Hove region).
This gap can be seen as the threshold energy
where these effects take place.
In the Van Hove scenario nodal quasiparticles are naturally
different from antinodal, the former being
strongly renormalized by scattering between the VH points.
Our proposal then is compatible with the opening of an
anisotropic gap, in fact it provides a physical
origin for the gap. In addition the Van Hove
quasiparticles have a lifetime compatible with
the marginal Fermi liquid.
It also agrees with
the recent proposal put forward in\cite{bontemps}
to explain the intraplane optical
conductivity and the c-axis optical conductivity
based on the difference between nodal and antinodal quasiparticles.

To illustrate these ideas we make a simple calculation. We compute
the Raman response in the absence of vertex corrections\cite{devereaux2}:
\begin{eqnarray}
\label{eq:raman}
\chi_\nu''({\bf q}=0,\Omega)&=&
\frac{2}{N}\sum_{\bf k}
\gamma_\nu^2({\bf k)}\int \frac{d\omega}{\pi}[f(\omega)-f(\omega+\Omega)]
\nonumber\\ {} &&G''({\bf k},\omega)
G''({\bf k},\omega+\Omega)
\end{eqnarray}
where $f(\omega)$ is the Fermi factor, $G''({\bf k},\omega)$ is the
imaginary part of the Green function of the electron,
$\gamma_{\nu}(\bf k)$ is the Raman vertex corresponding to
different channels $\nu=B_{1g},B_{2g}$ and the sum is all over the
Brillouin zone. The expressions for the unrenormalized Raman
vertices are $\gamma_{B_{1g}}{(\bf k)}=a (\cos k_x-\cos k_y)$ and
$\gamma_{B_{2g}}({\bf k})=b (\sin k_x\sin k_y)$\cite{devereaux2} where we
have absorbed constant factors in the $a$ and $b$ parameters.

The electron self-energy in
the Van Hove model has the form\cite{nuphysb97}:
\begin{equation}
\label{eq:sigmavh}
\Sigma_{VH}(\omega)=2\lambda\omega \log \frac{\omega}{\omega_c}-i\pi\lambda|\omega|,
\end{equation}
where
$\lambda$ is a dimensionless coupling constant which
is proportional to the onsite interaction, $\lambda \propto (U/t)^2$,
and $\omega_c$ is a cutoff. We assume
that this self-energy is a good approximation at
high temperature based on calculations using functional
renormalization group\cite{katanin}.

\begin{figure}[t]
  \begin{center}
       \epsfig{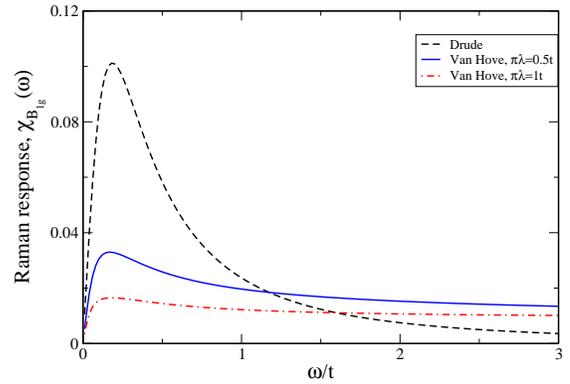}
\caption{Raman signal for $B_{1g}$ channels for the conventional
Drude behavior and for a system close to the
Van Hove filling. The calculations have been done for
$t'=-0.3t$ and $T=300K$ for different coupling constants. }
 \label{fig:vhnovh}
 \end{center}
\end{figure}

This self-energy is inserted in the spectral function
$A({\bf k},\omega)=-2 \rm Im G({\bf k},\omega+i\delta)$ where:
\begin{equation}
\label{eq:A}
A({\bf k},\omega)=\frac{-{\rm Im}\Sigma({\bf k},\omega)}
{(\omega+\mu-\epsilon_{\bf k}-
\rm Re\Sigma({\bf k},\omega))^2+{\rm Im}\Sigma({\bf k},\omega)^2},
\end{equation}
where
$\Sigma({\bf k},\omega)$ is the self-energy due to
either impurities or coupling to inelastic scattering
coming from Van Hove physics and
$\epsilon_{\bf k}=-2t(\cos k_x + \cos k_y)-4t'\cos k_x \cos k_y$.
The $A({\bf k},\omega)$ is
inserted in the Raman scattering given by Eq.~(\ref{eq:raman}).
The results for the $B_{1g}$ signal in units of $t$
are shown in Fig.\ref{fig:vhnovh}
where the standard Drude case (dashed curve) with a
constant scattering rate of $-{\rm Im}\Sigma=0.1t$ is shown for
comparison. To the imaginary part of the self energy
in the Van Hove model we have also added a constant
scattering rate turning up to be: $-{\rm Im}\Sigma=0.1+\pi\lambda|\omega|$.
We can identify the peak at a frequency
of twice the scattering rate $\omega=0.2t$ in the Drude curve.
For the Van Hove case we observe that
the peak is already strongly reduced for a
coupling constant of
$\pi\lambda=0.5t$, and can be completely washed out for bigger
coupling constant, $\pi\lambda=1t$,
showing the incoherent behavior.
In our argument the three curves in Fig.\ref{fig:vhnovh}
would  correspond to fillings
where the chemical potential is pinned at Van Hove filling
(Van Hove curve for $\pi\lambda=1t$),
fillings closer
to Van Hove filling
(Van Hove curve for $\pi\lambda=0.5t$),
and fillings above the Van Hove filling
(Drude curve).
We think that higher values of the coupling constant correspond
to decreasing dopings in cuprates being closer to the Mott
insulator.

In order to describe more quantitatively the metal-insulator transition
a full calculation must be performed varying
the doping around the Van Hove
filling\cite{prep}. We however believe
that the simple argument provided in this paper
may illustrate the physics of the problem.

Within our scheme we can also
address the temperature dependence of the Raman signal found
in \cite{devereaux1}.
The signal $B_{2g}$ corresponding to the nodal quasiparticles is
relatively doping independent and decreases with increasing temperature
showing a conventional behavior. In contrast, the signal $B_{1g}$
corresponding to antinodal quasiparticles varies strongly with
doping showing different temperature dependence for different
regions of dopings. In the
strongly overdoped region ($p\geq 0.22$) both $B_{1g}$ and $B_{2g}$
Raman signals decrease with
increasing temperature, a sign of
metallic behavior. For dopings in the range
$0.16\leq p \leq 0.20$ the $B_{1g}$ signal
becomes essentially temperature independent, and finally,
below optimal doping it increases with temperature. We
can understand this behavior qualitatively as due to the proximity of
the chemical potential to the
Van Hove singularity. Below and close to Van Hove filling the quasiparticles
can be
thermally excited, showing a semiconductor behavior.
In the strongly overdoped region we are far from the influence of
Van Hove and the behavior is the one expected in
a conventional metal. We do not have a clear understanding of the
intermediate region $0.16\leq p \leq 0.22$. In reference
\cite{devereaux1} it is argued that $p=0.22$ corresponds to a
quantum critical point where the anisotropy between the
$B_{1g}$ and $B_{2g}$ signals disappears. This feature goes beyond
the scope of the present study.

In all these qualitative arguments
one should bear in mind that the behavior seen in experiments is attached
to the fully interacting system and not to the
free one. Therefore we cannot determine where is the Van
Hove filling of the free theory.
Although in ARPES experiments\cite{vhexp} they can
observe an extended singularity,
the spectral function $A({\bf k}_{fixed},\omega)$ is extremely
broad and cannot be thought of as quasiparticles. That means that
any singularity in the density of states is washed out.
This
result is consistent with other experiments such as tunneling and
optical measurements where no singularity in the density of states
is observed.
This does not mean that the Van Hove singularity is not relevant for
the physics we are interested in, on the contrary, it is near the singularity
where one can expect a rich physics driven by the interactions of the enhanced
density of states
which encodes competing instabilities
and deformation of the Fermi surface\cite{vhscenario,metzbelweg}.
The situation is similar to the one related with the presence of a quantum
critical point  that can be inferred by nearby anomalies at finite
temperatures. In fact the Van Hove point is a quantum critical
point as the divergence of the density of states and the
various response functions at the critical
doping occurs at zero temperature.
We believe that the region of influence of the Van Hove singularity
is somewhere between optimal and overdoped doping because this is
the region where a dip in the Hall number is seen\cite{hall} and where the
renormalized Fermi surface seen by ARPES shows an extended
singularity\cite{vhexp}.

In summary, we propose that the metal-insulator transition
observed in Raman scattering in the antinodal region of BISCO
is driven by the incoherence of the transport set by the
strong scattering suffered by  antinodal quasiparticles
when the chemical potential is below or at the Van Hove singularity.

{\it Acknowledgments.} B.V. is supported by the -Postdoctoral
program at the University of Guelph. Funding from MCyT (Spain) through grant
MAT2002-0495-C02-01 is also acknowledged. One of us (B.V.) would like
to thank Thomas Devereaux and
Walter Metzner for valuable discussions. We are also very
grateful to
Paco Guinea for very useful conversations.

\end{document}